\documentclass[preprint]{aastex}
\usepackage[dvips]{epsfig,color}

\newcommand{\eg}{{\it e.g.,}}
\newcommand{\ie}{{\it i.e.,}}
\newcommand{\etal}{{\it et al.}}
\newcommand{\sersic}{S\'{e}rsic }

\newcommand{\psd}{\ensuremath{\rho/\sigma^{3}}}
\newcommand{\blo}{\ensuremath{\beta_{\rm lo}}}
\newcommand{\bhi}{\ensuremath{\beta_{\rm hi}}}

\newcommand{\ignore}[1]{\relax}

\shorttitle{Density Profiles II}
\shortauthors{Barnes \etal}

\begin{document}

\title{Density Profiles of Collisionless Equilibria. II. Anisotropic
Spherical Systems}
\author{Eric I. Barnes}
\affil{Department of Physics, University of Wisconsin---La Crosse,
La Crosse, WI 54601}
\email{barnes.eric@uwlax.edu}
\author{Liliya L. R. Williams}
\affil{Department of Astronomy, University of Minnesota, Minneapolis,
MN 55455}
\email{llrw@astro.umn.edu}
\author{Arif Babul\altaffilmark{1}}
\affil{Department of Physics \& Astronomy, University of Victoria, BC,
Canada}
\email{babul@uvic.ca}
\author{Julianne J. Dalcanton\altaffilmark{2}}
\affil{Department of Astronomy, University of Washington, Box 351580,
Seattle, WA 98195}
\email{jd@astro.washington.edu}

\altaffiltext{1}{Leverhulme Visiting Professor, Universities of Oxford
and Durham}
\altaffiltext{2}{Alfred P. Sloan Foundation Fellow}

\begin{abstract}

It has long been realized that dark matter halos formed in
cosmological N-body simulations are characterized by density profiles
$\rho(r)$ that, when suitably scaled, have similar shapes.
Additionally, combining the density and velocity dispersion profiles
$\sigma(r)$, each of which have decidedly nonpower-law shapes, leads
to quantity \psd\ that is a power-law in radius over 3 orders of
magnitude in radius.  Halos' velocity anisotropy profiles $\beta(r)$
vary from isotropic near the centers of halos to quite radially
anisotropic near the virial radius.  Finally, there appears to be a
nearly linear correlation between $\beta$ and the logarithmic density
slope $\gamma$ for a wide variety of halos.  This work is part of a
continuing investigation of the above interrelationships and their
origins using analytical and semi-analytical techniques.  Our finding
suggest that the nearly linear $\beta$--$\gamma$ relationship is not
just another expression of scale-free \psd\ behavior.  We also note
that simultaneously reproducing density and anisotropy profiles like
those found in simulations requires $\beta(r)$ and $\gamma(r)$ to have
similar shapes, leading to nearly linear $\beta$--$\gamma$
correlations.  This work suggests that the $\beta$--$\gamma$ and
power-law \psd\ relations have distinct physical origins.

\end{abstract}


\section{Introduction}\label{intro}

Cosmological N-body simulations are important tools that allow us to
reconstruct how gravitationally bound objects like galaxies and
clusters form and evolve in the universe.  The base-level simulations
involve Newtonian gravitational interactions between particles that
represent masses of dark matter exclusively.  These types of
simulations suppress the short-range interactions between masses,
making the dynamics collisionless, \ie\ particles move under the
influence of a global potential.  The consensus view of the results of
such simulations is that all bound structures have common density
structures determined by a few parameters.  The most often discussed
density distribution in the literature is the empircal
Navarro-Frenk-White \citep[][NFW]{nfw96,nfw97} function.  However,
many different functions have been invoked to describe the density
profiles of dark matter structures resulting from individual
simulations \citep[\eg][]{s68,m98,n04}.  While there is a considerable
amount of literature attempting to explain the universality of N-body
density profiles \citep{sw98,l00,n01,b05,lu06}, no widely accepted
explanation exists.

We are looking to gain a fundamental understanding of the physics
driving the universality of dark matter halos in collisionless
simulations.  Such insight may eventually provide a way of reconciling
the differences between simulated halos and the inferred properties of
halos of observed galaxies, such as the cusp--core controversy
\citep[\eg][]{bmr01,sgh05}. However, the results of the present paper
cannot be directly applied to observed galaxies and clusters because
the latter are affected by dissipational baryonic processes, which we
ignore in the present work; we concentrate exclusively on the effects
of gravity in collisionless systems.

In the past few years, two interesting relationships have been
uncovered that link various dynamical properties of halos formed in
simulations like those discussed above.  \citet{tn01} have found that
there is a scale-free relationship between a phase-space proxy and
radius; $\psd \propto r^{-\alpha}$, where $\rho$ is density, $\sigma$
is velocity dispersion, $r$ is radial distance, and $\alpha$ is a
constant for any given halo.  Interestingly, this relationship is not
unique to N-body results.  \citet{a05} have shown that halos created
semi-analytically also have power-law \psd\ behavior and speculate
that this feature is a result of violent relaxation.

The other relationship relates the density and velocity anisotropy
profiles of N-body halos \citep{hm04}.  Specifically, the curve
describing the logarithmic radial derivative of the logarithm of the
density has the same basic shape as the radial profile of a halo's
anisotropy.  As with the power-law behavior of \psd, there is no
obvious basis for this relationship, but together they signal that
there are unifying physical processes at work in halo formation.

This paper and its companion \citep[][hereafter Paper I]{b06} take
steps towards uncovering what these processes are.  In that work, we
found that the mechanical equilibrium of halos with velocity isotropy
is not the cause of power-law \psd\ profiles, as at least two
well-known equilibrium density profiles (Hernquist and King) do not
produce \psd\ power-laws.  However, \citet{n04} and \sersic density
distributions in mechanical equilibrium produce nearly scale-free
\psd\ profiles.  We have also investigated density profiles that
result from solving the isotropic Jeans equation with the constraint
that $\psd \propto r^{-\alpha}$.  In this restricted case, density
profiles are similar to those presented in \citet{n04} but are
reproduced better by \sersic models.

We continue and expand this work by again considering the density
profiles that result from solving the Jeans equation under the
assumption of scale-free \psd\ (motivated by N-body and
semi-analytical results), but now the velocity distributions will not
be restricted to be isotropic.  The details of solving the
``constrained'' Jeans equation are given in \S\ref{constrain}.  The
anisotropic Jeans equation admits a wide variety of solutions. We use
generic, but sensible, forms for our adopted anisotropy profiles.
These profiles and the solutions that result are discussed in detail
in \S\ref{constb} and \ref{varyb}.  Our main conclusion follows from
these sections; equilibrium density profiles replicate those seen in
simulations if the radial density and anisotropy profiles of any given
halo have similar shapes.  The degree of connectedness between the
\psd\ and density slope-anisotropy relations is considered in
Section~\ref{fundy}.  We end with a summary of our findings and our
conclusions.

\section{The Constrained Jeans Equation}\label{constrain}

The Jeans equation is a moment of the collisionless Boltzmann
equation.  Where the Boltzmann equation describes the evolution of
the full phase-space distribution function, which is not easily
obtained from simulations, let alone observations of real galaxies,
the Jeans equation relates observationally accessible quantities, like
density and velocity dispersion.  Mechanical equilibrium for a spherical
collisionless system with radially dependent anisotropy is determined
through the Jeans equation
\citep{j19,bt87},
\begin{equation}\label{sjeans}
\frac{d}{dr}\left[\frac{\rho(r) \sigma^2(r)}{3-2\beta(r)}\right]+
\frac{2\beta(r)}{3-2\beta(r)}\frac{\rho(r) \sigma^2(r)}{r}=
-G\rho(r)\frac{M(r)}{r^2},
\end{equation}
where $M(r)$ is the mass enclosed at radius $r$, $\rho$ is the
density, $\sigma$ is the total velocity dispersion, and $\beta$ is
the anisotropy.  The $3-2\beta(r)$ terms result from our choice to use
the total velocity dispersion rather than the $r$-component.  While we
do not directly calculate phase-space distribution functions here,
the density and velocity dispersion functions that will be discussed
are everywhere positive, implying that the corresponding
distribution functions are physically plausible.

One can reduce the number of functions in Equation~\ref{sjeans} by
assuming an ``equation of state'' that connects the density to the
total dispersion.  We use the empirically established relation
$\psd=(\rho_0/v_0^3)(r/r_0)^{-\alpha}$.
We note that while this connection is consistent with the
available numerical evidence \citep{tn01,dm05}, it is not the only
choice.  One could equally well choose to utilize the radial
dispersion, as in \citet{dm05}.  At this point in time, it is not
clear which choice is the most physically relevant.

Imposing this scale-free \psd\ constraint and changing to the
dimensionless variables $x \equiv r/r_0$ and $y \equiv \rho/\rho_0$,
we rewrite Equation~\ref{sjeans} as,
\begin{equation}\label{sjeans2}
-\frac{x^2}{y}\left[ \frac{d}{dx}\left(\frac{y^{5/3}x^{2\alpha/3}}
{3-2\beta(x)}\right)+\frac{2\beta(x)}{3-2\beta(x)}
y^{5/3}x^{2\alpha/3-1} \right]=BM(x),
\end{equation}
where $B=G/r_0 v_0^2$.  Differentiating this equation with respect to
$x$ gives us,
\begin{equation}\label{sjeans3}
\frac{d}{dx}\left[-\frac{x^2}{y}\left\{
\frac{d}{dx}\left(\frac{y^{5/3}x^{2\alpha/3}}
{3-2\beta(x)}\right)+\frac{2\beta(x)}{3-2\beta(x)}
y^{5/3}x^{2\alpha/3-1} \right\} \right]=Cyx^2,
\end{equation}
where $C=4\pi\rho_0 r_0^2/v_0^2$.  This expression is an extension of
equation (8) in Paper I.  We eliminate the constant $C$ by
solving for $y$, differentiating with respect to $x$ again, and
grouping like terms.  The resulting constrained Jeans equation is,
\begin{eqnarray}\label{cjeans}
(2\alpha+\gamma-6)(\frac{2}{3}(\alpha-\gamma)+1)(2\alpha-5\gamma)&=&
15\gamma''+3\gamma'(8\alpha-5\gamma+4\beta+
12\beta\delta b_1 -5)\\ \nonumber
& &-3\delta[b_1 (4\alpha^2+\gamma^2-8\alpha\gamma
+8\alpha+7\gamma-15)]\\ \nonumber
& &-3\delta^2[6b_1 b_2 (\alpha-\gamma+1)]\\ \nonumber
& &-3\delta^3[b_3 (54\beta +144\beta^2 +24\beta^3)]\\ \nonumber
& &-3\delta'[6b_1 (\alpha-\gamma+1)+9b_1 b_2 \delta]
-3\delta''[3b_1]
\end{eqnarray}
In this notation, $\gamma=\gamma(x)=-d\ln{y}/d\ln{x}$ is
the logarithmic density slope,
$\delta=\delta(x)=d\ln{\beta}/d\ln{x}$, $b_1=2\beta/(3-2\beta)$,
$b_2=(3+2\beta)/(3-2\beta)$, $b_3=(3-2\beta)^{-3}$, and the primes
indicate derivatives with respect to $\ln{x}$.

As discussed previously, the parameter $\alpha$ is a constant, and we
choose to investigate the following values; $\alpha=1.875$,
$35/18=1.94\bar{4}$, 1.975.  These values have been chosen based on
the findings of \citet{a05}.  For isotropic systems, $\alpha=35/18$
represents a critical point for solutions of the constrained Jeans
equation.  Solutions with $\alpha > 35/18$ have $\gamma$ values that
asymptote to finite values, and $\alpha=1.975$ is our example of such
a case.  In terms of density, these solutions approach power-law
density profiles at large radius.  In the other case, $\gamma$ values
increase indefinitely with radius, implying density profiles that
continually steepen farther from the center.  Our value of
$\alpha=1.875$ (which is also the value found by \citet{tn01})
represents this class of solutions.

While aesthetically unappealing in the extreme, Equation~\ref{cjeans}
can be straightforwardly solved numerically for $\gamma$ over a range
of $\ln{x}$ once $\alpha$ and $\beta(x)$ have been specified
(different choices of $\beta(x)$ are considered in \S\S \ref{constb},
\ref{plb}, and \ref{tanb}).  Our approach to the solution sets
$\gamma=2$ at the scalelength $r_0$ ($x=1$).  In effect, we define the
scalelength of a halo to be the radius where the density slope has the
isothermal value.  We further define $x_{\rm{vir}}=10$, by analogy
with $c=10$ NFW profiles [note that this does not mean that
$\bar{\rho}(x<10)=200\rho_{\rm{crit}}$].  The remaining initial
condition is the value of $\gamma'$ at $r_0$.  We leave this as a free
parameter to be determined by minimizing the differences between the
solution $\gamma$-profile and those of the NFW,
\begin{eqnarray}\label{nfwdef}
\rho(x)/\rho_0 & = & x^{-1} (1+x)^{-2} \nonumber \\
\gamma(x) & = & (1+3x)/(1+x)
\end{eqnarray}
and
\citet[][N04]{n04}
\begin{eqnarray}\label{n04def}
\log{[\rho(x)/\rho_0]} & = & (-2/\mu)(x^{\mu}-1) \nonumber \\
\gamma(x) & = & 2x^{\mu}
\end{eqnarray}
density profiles.  In particular, for the NFW profile $\gamma'(1)=0.5$
and for the N04 profile $\gamma'(1)=2\mu$, where $\mu=0.17$ (the
best-fit value from N04).  We have chosen these profiles because of
their ability to empirically describe the collisionless halos that
form in a wide variety of numerical simulations as well as their
relative simplicity, but any other profile could be chosen and
analyzed in the same way.

\section{Constant Anisotropy Solutions}\label{constb}

The simplest type of anisotropy distribution is a constant nonzero
value.  In this case, equation~\ref{cjeans} simplifies considerably as
$\delta$ and its derivatives disappear.  Our work in Paper I deals
with the special case of $\beta=0$.  Here, we extend to nonzero, but
constant, anisotropies ($-0.2\le \beta \le 1.0$).  The relevant
constrained Jeans equation is,
\begin{equation}\label{cjeans2}
(2\alpha+\gamma-6)(\frac{2}{3}(\alpha-\gamma)+1)(2\alpha-5\gamma)=
15\gamma''+3\gamma'(8\alpha-5\gamma+4\beta-5).
\end{equation}

Solutions to this equation extend and agree with our earlier work.  In
general, solutions to the constant anisotropy constrained Jeans
equation do not resemble NFW profiles for any tested combination of
constant anisotropy value and $\alpha$ value.  On the other hand, the
constrained Jeans equation solutions can approximate N04 profiles, at
least for isotropic and slightly tangentially anisotropic velocity
dispersion distributions.  For the solutions that best approximate N04
profiles, the anisotropy value decreases as the $\alpha$ value
increases.  Figure~\ref{gp1} shows $\gamma$ profiles for
$\alpha=1.875, \beta=0.1$ (panel a), $\alpha=35/18=1.94\bar{4},
\beta=-0.1$ (panel b), and $\alpha=1.975, \beta=-0.2$ (panel c).  Even
mild radial anisotropy ($\beta \ga 0.3$) drives the solutions to have
fairly constant $\gamma$ profiles, independent of $\alpha$.  Panel d
of Figure~\ref{gp1} shows the $\gamma$ profile for $\alpha=1.875$,
$\beta=0.5$.  We note that the special value $\alpha=35/18$ is
obtained analytically only for $\beta=0$ \citep{a05}.  For nonzero
constant $\beta$, the special value of $\alpha$ can be found by
setting the $\gamma'$ term in Equation~\ref{cjeans2} equal to zero and
utilizing the relationship for power-law density profiles,
$\gamma_{\rm pl}=6-2\alpha$.  The result is that $\alpha_{\rm
special}=35/18-2\beta/9$ \citep[see also][]{dm05}.

\section{Solutions for Anisotropy Distributions}\label{varyb}

On relaxing the constant anisotropy requirement, we are immediately
faced with a decision.  What form should the anisotropy distribution
$\beta(x)$ take?  We have chosen two flexible functions to
investigate; one is a simple power-law while the other is based on the
hyperbolic tangent function.  For both functions, we choose boundary
anisotropy values and we have parameters that affect the rates of
change between these boundary values.  The functions are similar in
that they both monotonically increase with $\ln{x}$, following the
stability argument given in \citet{ma85}.  As the next two subsections
will show, the important difference between the two functions is that
the power-law has a positive second derivative everywhere while the
hyperbolic tangent has an inflection point.

Specifically, the power-law function is given by,
\begin{equation}
\beta(x)=\blo+(\bhi-\blo)
\left[\frac{\ln{x}+\ln{x_{\rm bound}}}
{\ln{x_{\rm vir}}+\ln{x_{\rm bound}}}\right]^{\eta},
\end{equation}
where \blo\ is the anisotropy value as $\ln{x}\rightarrow -\ln{x_{\rm
bound}}$, \bhi\ is the anisotropy value at the scaled virial radius
$x_{\rm vir}$, $\eta>0$ controls the shape of the distribution, and
the range of integration is determined by $\ln{x_{\rm bound}}=25$.  We
fix the scaled virial radius at the value for an NFW halo with a
concentration index of 10, $x_{\rm vir}=10$.  While our solutions
cover large ranges of $x$, we will be interested primarily in the
several orders of magnitude surrounding the scalelength.

The other distribution that we study will be referred to as the tanh
anisotropy profile.  The expression for this distribution is,
\begin{equation}
\beta(x)=\frac{1}{2}(\bhi-\blo)[1+\tanh{(s\ln{x})}]+\blo,
\end{equation}
where again \blo\ is the anisotropy as $\ln{x}\rightarrow -\ln{x_{\rm
bound}}$, \bhi\ is now the anisotropy as $\ln{x}\rightarrow \ln{x_{\rm
bound}}$, and $s$ determines the rate of change between these boundary
values.  Now, $\tanh{z}=(e^z-e^{-z})/(e^z+e^{-z})$ and
$1+\tanh{z}=2e^z/(e^z+e^{-z})$.  If $z=s\ln{x}=\ln{x^s}$, then
$1+\tanh{z}=2x^s/(x^s+x^{-s})=2x^{2s}/(x^{2s}+1)$.  The tanh form is
thus a generalization of commonly discussed anisotropy functions;
$\beta \propto r^2/(r^2+r_a^2)$ (Osipkov-Merritt) for $s=1$, and
$\beta \propto r/(r+a)$ \citep[\eg][]{ml05} for $s=1/2$.  The tanh
form we adopt allows for more flexibility in the anisotropy
profile.

Our approach to solving Equation~\ref{cjeans} with these anisotropy
distributions is as follows.  First, choose a profile type to fit to;
either the NFW or N04 $\gamma$ distribution.  Next, choose pairs of
(\blo, \bhi) values.  With the boundary anisotropy values fixed, we
are left with one free parameter in each $\beta$ profile.  For each
set of (\blo, \bhi), we find the $\eta$ or $s$ that minimizes the
average least-squares difference between the $\gamma$ profile that
results from solving Equation~\ref{cjeans} and the chosen $\gamma$
distribution over the range $10^{-3} x_{\rm vir}$ and $x_{\rm vir}$, a
range over which N-body simulations produce meaningful results.  This
procedure provides us with the best approximation to either the NFW or
N04 profiles given a set of boundary anisotropy values.  Note that we
do not use the least-squares values for hypothesis testing or for
making statistical comparisons between the various anisotropy forms,
but rather to find the form of the anisotropy profile that best
reproduces NFW/N04 density profiles, while solving the Jeans equation.

\subsection{Power-law Anisotropy Solutions}\label{plb}

For the power-law anisotropy solutions, we have investigated a
parameter space with $0.0 \le \blo \le 0.9$ and $0.1 \le \bhi \le
1.0$.  Points where $\blo = \bhi$ have not been included as they are
identical to the constant anisotropy models (\S \ref{constb}).  Points
with $\blo,\bhi<0$ have been excluded due to an instability in the
fitting routine for these models.  We have also investigated the
regions of parameter space where $\blo > \bhi$, but those fits are
uniformly uninteresting as the solutions do not resemble any sort of
physically relevant density profile.  The solutions which best match
NFW and N04 profiles all have $\blo = 0$.  Solutions that are matched
to NFW profiles tend to have smaller best-fit \bhi\ values than those
matched to N04 profiles.  The best-fit solutions are discussed in
detail below.

The best-fit $\gamma$ distributions with power-law anisotropy profiles
are shown in Figure~\ref{plbnfw}, where the solutions (solid lines)
are matched to the NFW profile (dashed lines).  Dash-dotted lines
illustrate N04 $\gamma$ profiles.  Dotted lines show the anisotropy
profile from \citet{ml05}.  Panels a and b show the $\gamma$ and
$\beta$ profiles calculated with $\alpha=1.875$ that best match the
appropriate profile, panels (c, d) and (e, f) illustrate the same
distributions, but for $\alpha=35/18$ and $\alpha=1.975$,
respectively.  Note that the solutions for the power-law anisotropy
profile never resemble the NFW profile; \ie\ the best-fits are not
good fits.

Figure~\ref{plbn04} illustrates the same relations, but the solutions
are matched to the N04 profile.  The line styles are the same as in
Figure~\ref{plbnfw}.  The N04 profile is reproduced with much greater
accuracy than the NFW profile, but the match becomes substantially
worse as $\alpha$ increases.  Note that for the solutions that match
the N04 profiles, the best-fit anisotropy profile has an isotropic
distribution near the center and a highly radial anisotropic character
($0.8 \la \beta \la 1.0$) near the virial radius.  This character of
anisotropy profiles has been noted previously in N-body simulations
\citep[\eg][]{va82,cl96}.  It has also been suggested that such
behavior is evidence that an orbital stability process is at work in
determining the density structure of equilibrium halos \citep{b05}.

In numerous N-body simulations, a nearly linear relationship has been
found between the anisotropy and density-slope profiles
\citep{hm04,hm06}.  It is important to note that no one simulation
displays an exactly linear $\beta$--$\gamma$ profile.  Over the range
of $\gamma$ values our solutions cover ($1\la \gamma \la 3$),
individual simulation $\beta$--$\gamma$ curves are nearly linear
indicating that the anisotropy and density slope distributions have
similar shapes.  The variations present in Figure 2 of \citet{hm06}
appear to originate from differing slopes and intercepts for different
simulations.

In Figures~\ref{plhmnfw} and \ref{plhmn04}, we present the $\beta$ vs.
$\gamma$ profiles (solid lines) obtained from the solutions shown in
Figures~\ref{plbnfw} and \ref{plbn04} and compare them to the
relations found in N-body simulations.  The dashed lines represent the
mean trend of the simulation results and the dash-dotted lines show
the extent of variations in those results.  Of the six curves in
Figures~\ref{plhmnfw} and \ref{plhmn04}, the best qualitative match to
the near-linear relationship found in Figure 2 of \citet{hm06} is
given by the N04-like profile with $\alpha=1.875$, but other values
$1.8 \lesssim \alpha \lesssim 1.9$ would also give similar results
(Figure~\ref{plhmn04}a).  One difference is that our relation is
``steeper''; we reach total radial anisotropy at $\gamma=3$, while the
mean simulation results reach $\beta \approx 0.5$ at $\gamma=3$.
Solving the constrained Jeans equation for N04-like profiles with
$\alpha \gtrsim 1.9$ or for NFW-like profiles does not produce
near-linear $\beta$--$\gamma$ relations, but the solutions are at
least consistent with the mean trend within the variations seen in
simulations.

\subsection{Tanh Anisotropy Solutions}\label{tanb}

We now turn to the hyperbolic tangent (tanh) anisotropy distribution
solutions.  The parameter space for the tanh profile is slightly
different than for the power-law anisotropy distribution; $-0.2 \le
\blo \le 0.9$ and $0.3 \le \bhi \le 1.0$.  Other combinations of \blo
and \bhi\ yield unrealistic density profiles.  As in the power-law
case, the \blo values for the solutions that best match both NFW and
N04 profiles are small, $\blo \le 0.1$.  The \bhi\ values for the NFW
matched solutions are larger than those for the N04 matched solutions.
The specifics of the best-fit solutions are given below.

The solutions that best approximate the NFW profile have anisotropy
profiles that vary from isotropic near the center to mildly radially
anisotropic ($\beta \ga 0.5$) at the virial radius.  The best-fit
solutions' $\gamma$ and $\beta$ profiles are shown in
Figure~\ref{tbnfw}.  Again, the line styles are the same as in
Figure~\ref{plbnfw}.  Unlike in the power-law $\beta$ case, it is easy
to see that these solutions approximate the asymptotic NFW behavior
quite well.

Another difference from the power-law case is the behavior of the
solutions that resemble N04 profiles.  These solutions tend to have
fairly constant anistropy distributions that are nearly isotropic
($\beta \la 0.3$).  This is most evident when we look at the best-fit
solution $\gamma$ and $\beta$ profiles in Figure~\ref{tbn04}.  The N04
$\gamma$ profile is well-fit by the solution, but only when the
anisotropy profile is nearly constant near zero.  This agrees with the
work presented in Paper I where a $\beta=0$ profile produces solutions
that are good approximations to N04 profiles.

Figures~\ref{thmnfw} and \ref{thmn04} show the $\beta$ vs. $\gamma$
curves for the NFW-like and N04-like solutions, respectively.  The
curve for the NFW-like solution with $\alpha=1.875$
(Figure~\ref{thmnfw}a) is remarkably linear, again indicating the
similarity of the $\beta$ and $\gamma$ distributions.  In fact, each
of the curves in Figure~\ref{thmnfw} have substantial regions of
linearity.  This behavior makes them all qualitatively similar to the
results of \citet{hm06} (marked by the dashed and dash-dotted lines as
before).  However, the most linear relation (panel a with
$\alpha=1.875$) is also the one that disagrees most strongly with the
simulation results.  The remaining profiles ($\alpha=1.94\bar{4}$ and
$\alpha=1.975$) actually show a decent quantitative match to the mean
simulation trend.  Figure~\ref{thmn04} shows that the N04-like density
profile that is most linear (again $\alpha=1.875$) is a poor
quantitative match to simulation results.  Overall, the N04-like
density profiles produce nearly horizontal $\beta$--$\gamma$ profiles,
qualitatively dissimilar to the profiles from simulations.  However,
we note that profiles created with larger ($\ga 1.94\bar{4}$) $\alpha$
values are quantitatively consistent with simulations.

\section{Relating Scale-free \psd\ and $\beta$ vs. 
$\gamma$}\label{fundy}

As these results have been found by assuming the power-law nature of
\psd, one could ask whether or not the $\beta$ vs. $\gamma$ behavior
is independent of scale-free \psd.  Since there is no known
theoretical motivation for either of these relationships, we attack
this question with different tactics.  

The extended secondary infall model (ESIM) \citep{w04,a05} is a
semi-analytical halo formation scenario based on the work of
\citet{rg87}.  In ESIM, spherical halos form by accreting shells of
material that have decoupled from Hubble expansion.  These halos exist
at overdense locations and their formation includes effects from
secondary perturbations that can be completely specified and
controlled.  As shells collapse, the shells' energies change in
response to the continually changing potential. In fact, the ESIM halo
collapse process is best described as violent relaxation because
direct two-body effects do not take place (there are no shell-shell
interactions), while the shells are allowed to exchange energy by
interaction with the global potential.  In general, ESIM halos have
power-law density distributions over the radial ranges in which N-body
simulations show decidedly non-power law behavior, \eg\ where NFW
profiles change from $\rho \propto r^{-1}$ to $\rho \propto r^{-3}$.

Standard ESIM halos have scale-free \psd\ \citep{a05,b05}, but what do
their $\beta$ vs. $\gamma$ curves look like?  Panel a of
Figure~\ref{esim} represents the $\beta$ and $\gamma$ values taken
from a standard ESIM halo, and the locus bears no resemblence to the
linear relation found in N-body simulations.  This implies that the
$\beta$--$\gamma$ relationship is not just a manifestation of a
scale-free \psd.  It is certainly possible that the process
responsible for creating scale-free \psd\ is not fully expressed in
standard ESIM simulations, leading to a different $\beta$--$\gamma$
relationship.  However, we suggest that the near-linear relation
between $\beta$ and $\gamma$ supports an earlier argument that one
generic outcome of mostly-radial collapses is an instability that
produces velocity distributions with nearly isotropic cores and more
radially anisotropic envelopes \citep{b05}.  Due to the
semi-analytical nature of the ESIM formalism, this instability cannot
develop in ESIM halos.  However, by manipulating the secondary
perturbations present in ESIM halo formation, one can induce this type
of anisotropy profile in ESIM halos.  Doing so radically changes the
associated density profile from the single power-law of the standard
halo to an NFW-like form \citep{b05}.  These changes strongly affect
the overall shape of the $\beta$ vs.  $\gamma$ plot.  Panel b of
Figure~\ref{esim} illustrates the $\beta$ vs.  $\gamma$ relationship
for an NFW-like ESIM halo.  While the points certainly do not fall on
a line, they more closely approximate the N-body simulation results.
The implication is that the physics driving the density profile shape
is also important to establishing the linear $\beta$--$\gamma$
relationship.

\section{Summary \& Conclusions}

It is well-established empirically that \psd\ is a power-law in radius
for virialized, collisionless halos \citep{tn01,a05,dm05}.  We exploit
this relationship and transform the usual Jeans equation into the
constrained Jeans equation, solely in terms of the logarithmic density
slope $\gamma$ and the anisotropy $\beta$.  Doing this allows us to
solve directly for $\gamma$ once an anisotropy distribution has been
chosen.

In an earlier paper, \citet{b06}, we discussed the solutions of this
constrained Jeans equation when the anisotropy is assumed to be zero
everywhere.  Here, we have extended our earlier work to include
constant, but nonzero, anisotropies as well as anisotropy
distributions.  In particular, we have looked at two flexible,
but distinct, types of anisotropy distributions; a power-law and a
hyperbolic tangent.  We find that, in general, the solutions of the
constrained Jeans equation that most resemble the empirical NFW and
N04 density profiles have $\blo \approx 0$ and $\bhi \ga 0.5$.
Anisotropy must be an increasing function of radius in halos like the
ones produced in N-body simulations.

A central conclusion of our present work is that there is a distinct
similarity between the shapes of radial density and anisotropy
profiles.  Density profiles that best match the NFW form are found
when the anisotropy profile is of the tanh form (right panels of
Figure~\ref{tbnfw}), which contains the Osipkov-Merritt and
Mamon-Lokas forms as special cases.  N04-like solutions are found with
either profile (right panels of Figures~\ref{plbn04} and \ref{tbn04}),
but with the tanh $\beta$ profile, the anisotropy must be roughly
constant over the range $10^{-3} r_{\rm vir} < r < r_{\rm vir}$
(Figure~\ref{tbn04}).

Another aspect of this profile similarity is given by the $\beta$ vs.
$\gamma$ plots (Figures~\ref{plhmnfw}, \ref{plhmn04}, \ref{thmnfw},
and \ref{thmn04}).  The combination of $\beta$ and $\gamma$ profiles
that produce the most linear curves \citep[as suggested by the results
of N-body simulations;][]{hm04,hm06}, are, of course, those profiles
that resemble each other most closely.  The inflection point in the
tanh anisotropy distribution also appears in the NFW density slope
distribution (right panels of Figure~\ref{tbnfw} and
Figure~\ref{thmnfw}).  In other words, when this second-derivative
symmetry is lost, as with NFW-like density and power-law anisotropy
profiles (right panels of Figure~\ref{plbnfw}), so is the near-linear
correlation between $\beta$ and $\gamma$.  However, we note that the
most linear $\beta$--$\gamma$ relations found here are also those that
are the most quantitatively different from the results of simulations,
in that the solid red lines in Figures~\ref{plhmn04}a and
\ref{thmnfw}a extend outside the simulation uncertainty range.  This
suggests that the relationship between $\beta$ and $\gamma$ for any
single halo is not strictly linear.

Utilizing the results of semi-analytic models of halo formation, we
argue that the $\beta$--$\gamma$ relationship is not just a
manifestation of a scale-free \psd.  Rather, a near-linear relation
between $\beta$ and $\gamma$ supports an earlier argument regarding
the nature of halo density profiles \citep{b05}.  Velocity
distributions with nearly isotropic cores and more radially
anisotropic envelopes result from radial collapses.  This form of
anistropy distribution necessarily creates a density profile that is
less steeply rising in the central regions, since there are fewer
radial orbits that plunge close to the center and can support a strong
cusp.  Likewise, more tangential orbits in the outer regions would
increase the density since the average time an orbit spends far from
the center would increase.  At least in a qualitative sense, this
picture supports the observed similarities between profile shapes;
more isotropy (smaller $\beta$) implies less steeply varying density
(smaller $\gamma$) and vice versa.

The outcome of this work is that there appear to be at least two
physical processes at work in the formation of collisionless dark
matter halos in N-body simulations.  The evidence presented suggests
that one process creates the scale-free \psd\ distribution and must be
a fairly generic collapse process as the \psd\ behavior is seen in the
more restrictive ESIM halos.  As a consequence, another process links
the anisotropy and density profiles in a unique, nearly linear way.

\acknowledgments
We thank an anonymous referee for several helpful suggestions.  This
work has been supported by NSF grant AST-0307604.  Research support
for AB comes from the Natural Sciences and Engineering Research
Council (Canada) through the Discovery grant program.  AB would also
like to acknowledge support from the Leverhulme Trust (UK) in the form
of the Leverhulme Visiting Professorship at the Universities of Oxford
and Durham.  JJD was partially supported through the Alfred P. Sloan
Foundation.

\begin{figure}
\plotone{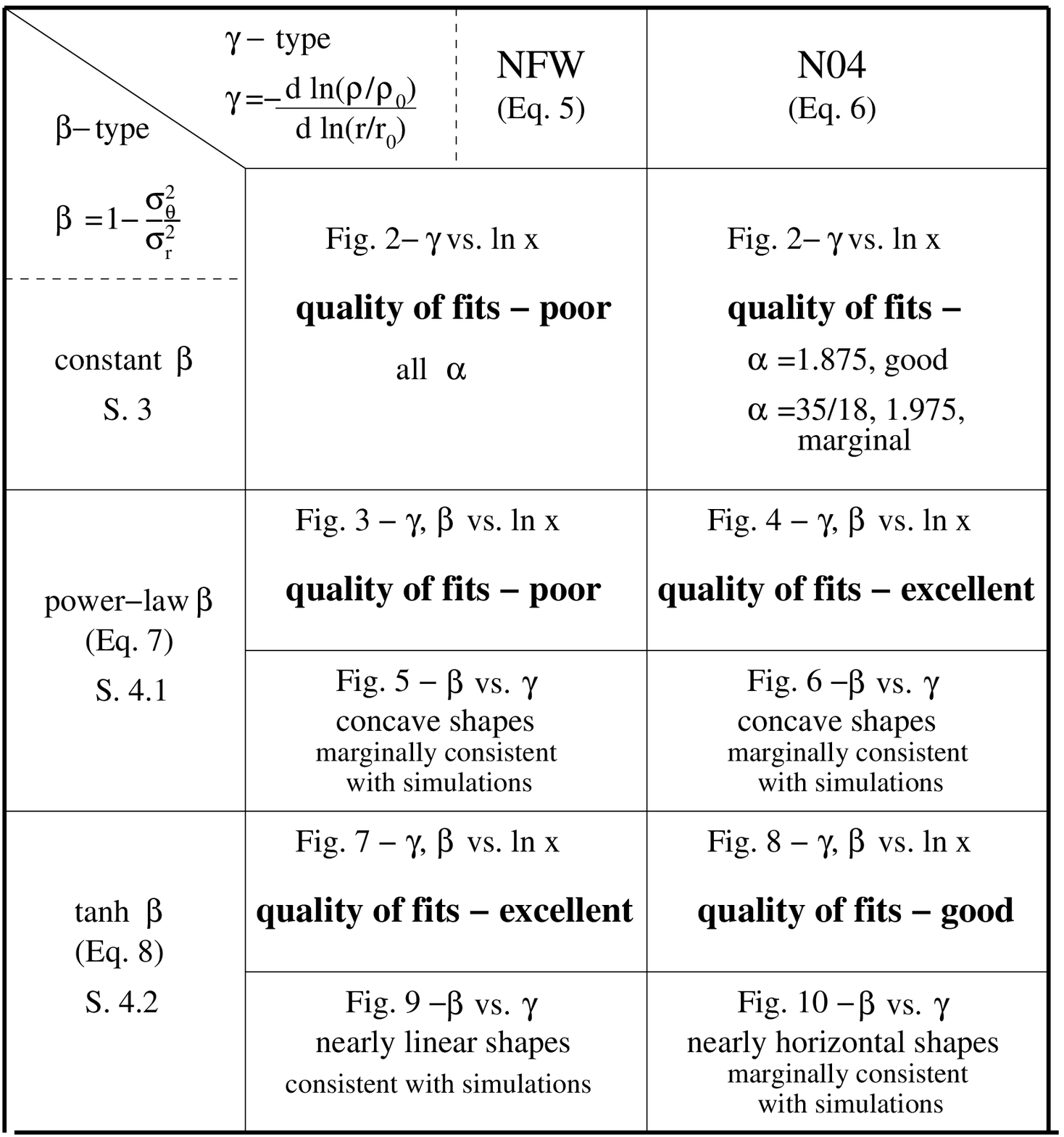}
\figcaption{An overview of the models and results presented in this
paper.  The columns refer to the two types of density profiles
discussed here; Navarro-Frenk-White \citep[][NFW]{nfw96,nfw97} and
\citet[][N04]{n04}.  These profiles are described in detail in
\S \ref{constrain}.  The rows delineate the three types of anisotropy
profiles; constant (\S \ref{constb}), power-law (\S \ref{plb}), and
hyperbolic tangent (\S \ref{tanb}).  This table is intended to guide
the reader through and summarize the various figures and key results.
\label{table}}
\end{figure}

\begin{figure}
\plotone{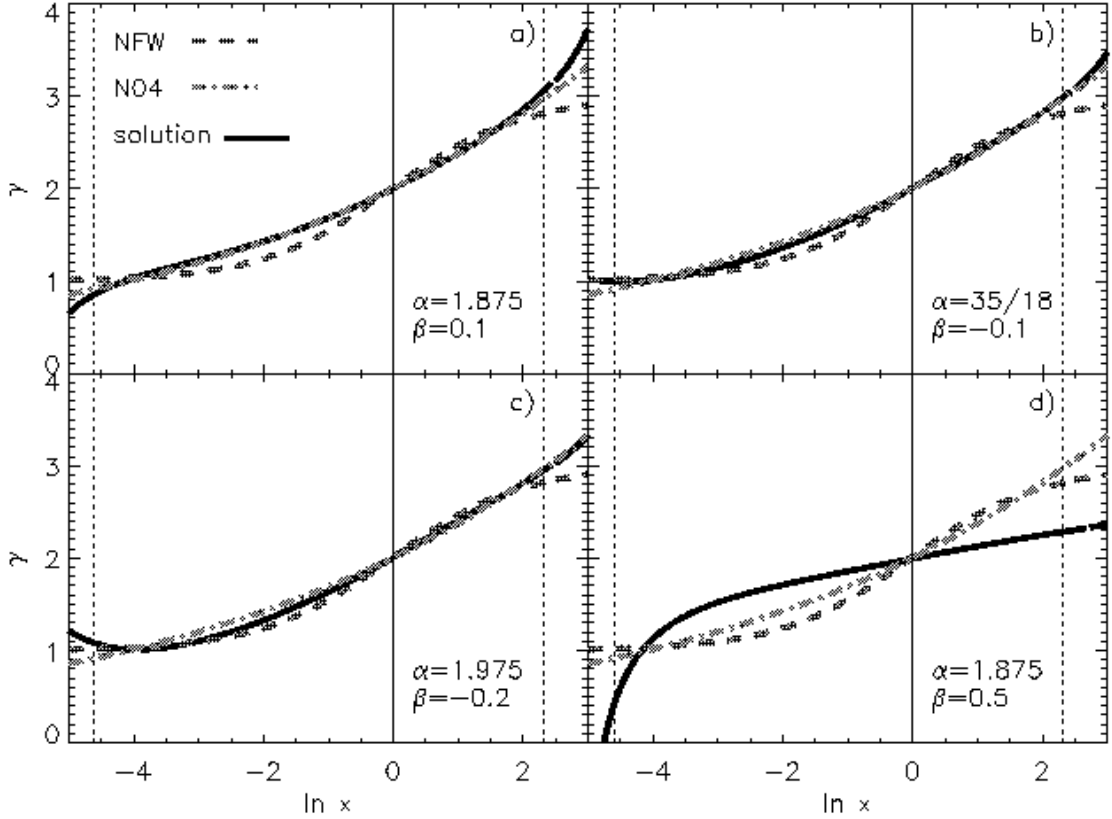}
\figcaption{The $\gamma=-d\ln{(\rho/\rho_0)}/d\ln{(r/r_0)}$
distributions (solid red lines) that result from solving the
constrained Jeans equation (Equation~\ref{cjeans2}) with constant
anisotropy $\beta$ values compared with N04 (Equation~\ref{n04def},
dash-dotted green lines) and NFW (Equation~\ref{nfwdef}, dashed blue
lines) $\gamma$ profiles.  Panels a-c show the solutions that best-fit
the N04 profile; no combinations of $\beta$ and $\alpha$ produced
solutions that approximate NFW profiles.  Panel a shows the solution
with $\alpha=1.875, \beta=0.1$.  The solution in panel b has
$\alpha=35/18, \beta=-0.1$, and panel c has $\alpha=1.975,
\beta=-0.2$.  The solution in panel d has $\alpha=1.875, \beta=0.5$.
It is included to illustrate the behavior of radially anisotropic
systems.  All solutions, regardless of $\alpha$ values, with $\beta
\ga 0.3$ resemble this solution and provide poor approximations to any
empirical $\gamma$ distribution.  See \S \ref{constb} for details.
\label{gp1}}
\end{figure}

\begin{figure}
\plotone{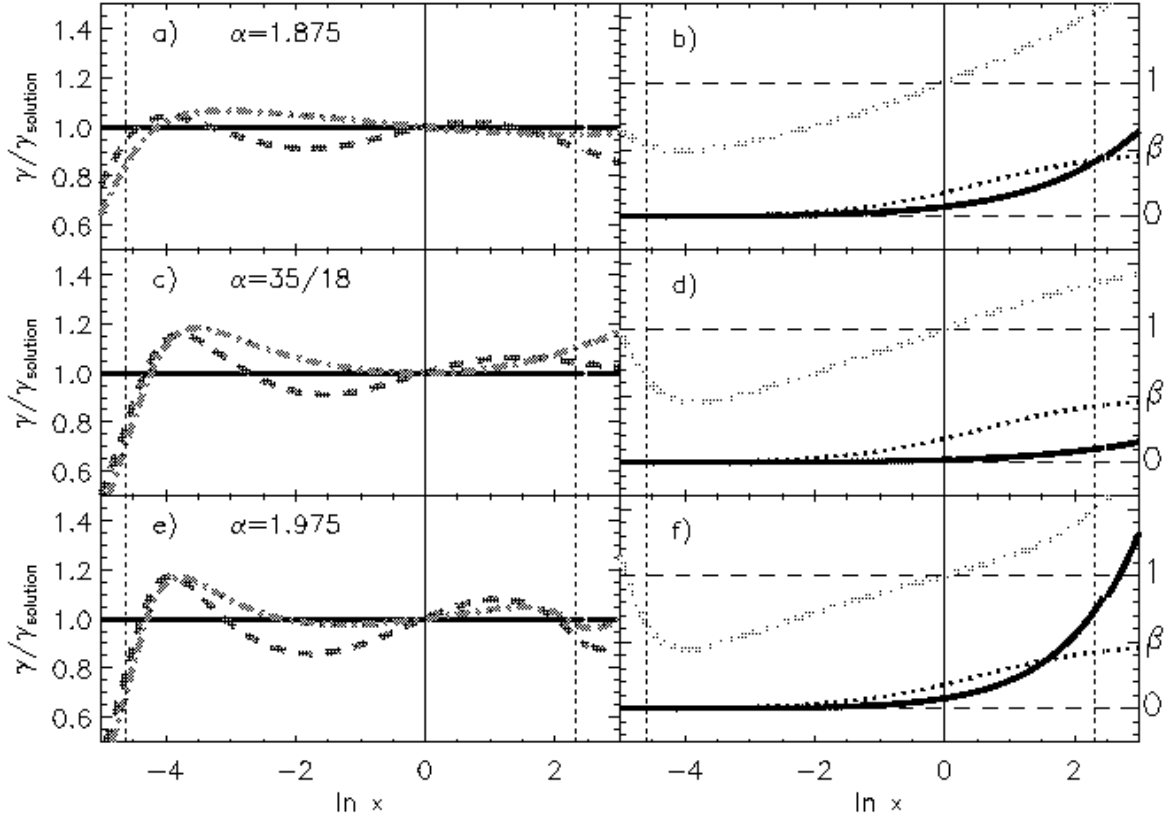}
\figcaption{With power-law $\beta(x)$, panels a, c, and e show the
calculated $\gamma$ profiles (solid red lines) that best match the NFW
$\gamma$ profile (dashed blue lines) for $\alpha=1.875,\mbox{
}35/18,\mbox{ and }1.975$, respectively.  To magnify the variations,
each $\gamma$ profile has been normalized by the solution.  The
dot-dashed green lines show the N04 $\gamma$ profile.  Panels b, d,
and f illustrate the anisotropy distributions used to calculate the
solutions (solid red lines).  The orange triple-dot-dashed lines are
scaled versions of the solution $\gamma$ profile and are included to
compare the shapes of the $\beta$ and $\gamma$ profiles.  The thick
dotted lines represent an anisotropy distribution commonly used to
describe the results of N-body simulations \citep[][Equation
60]{ml05}.  The vertical dotted lines mark the boundaries of the
fitting region, $x=0.01$ and $x=10$.  The solutions never
well-approximate the NFW profile.  In fact, for $\alpha=1.875$, the
solution is much more like an N04 profile.  Note that an NFW $\gamma$
profile has an inflection point at $\ln{x}=0$, whereas this $\beta$
profile does not.  See \S \ref{plb} for details.
\label{plbnfw}}
\end{figure}

\begin{figure}
\plotone{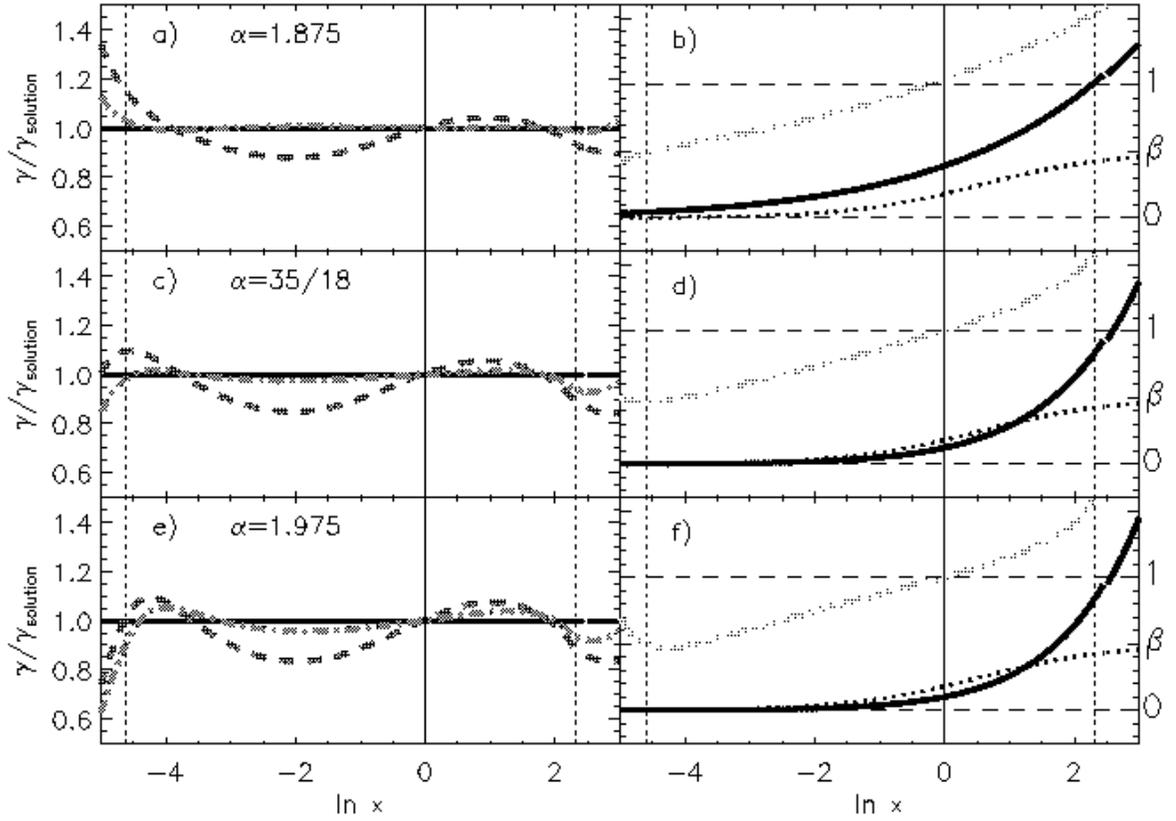}
\figcaption{With power-law $\beta(x)$, panels a, c, and e show the
calculated $\gamma$ profiles that best match the N04 $\gamma$ profile
for $\alpha=1.875,\mbox{ }35/18,\mbox{ and }1.975$, respectively.
Panels b, d, and f illustrate the anisotropy distributions used to
calculate the solutions.  The line styles are the same as in
Figure~\ref{plbnfw}.  The solutions match the designated N04 profile
reasonably well, in contrast with the results shown in
Figure~\ref{plbnfw}.
\label{plbn04}}
\end{figure}

\begin{figure}
\plotone{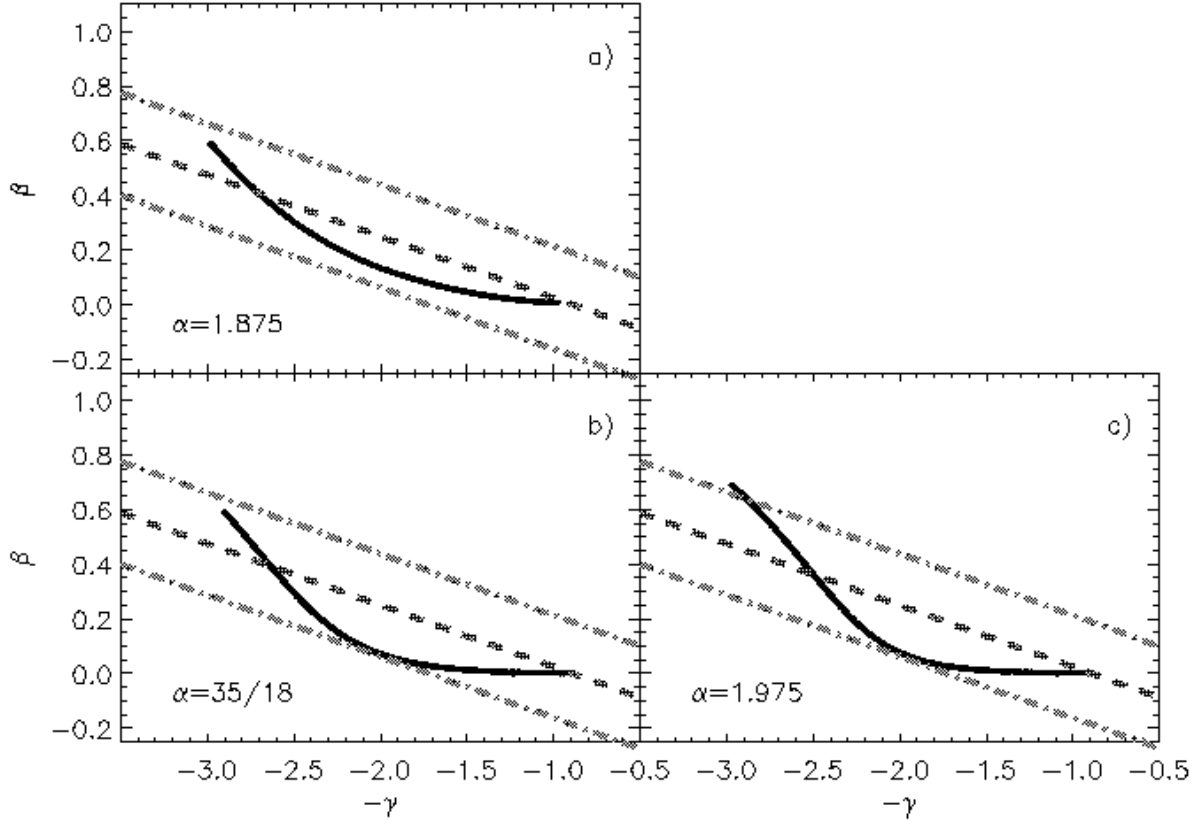}
\figcaption{Correlations between power-law velocity anisotropy $\beta$
and logarithmic density slope $-\gamma$ profiles of constrained Jeans
equation solutions for $10^{-3}x_{\rm vir} \le x \le x_{\rm vir}$.
Panels a, b, and c represent the best-fit NFW solutions with
$\alpha=1.875,\mbox{ } 35/18, \mbox{ and } 1.975$, respectively.  The
dashed blue line represents the mean trend found in simulation results
\citep{hm06} and the dash-dotted green lines mark the extent of
variations from those simulations.  While the solutions are
quantitatively consistent with the simulations, their shapes
(particularly the concavity) are unlike the results of simulations
over similar ranges of $\gamma$ values.
\label{plhmnfw}}
\end{figure}

\begin{figure}
\plotone{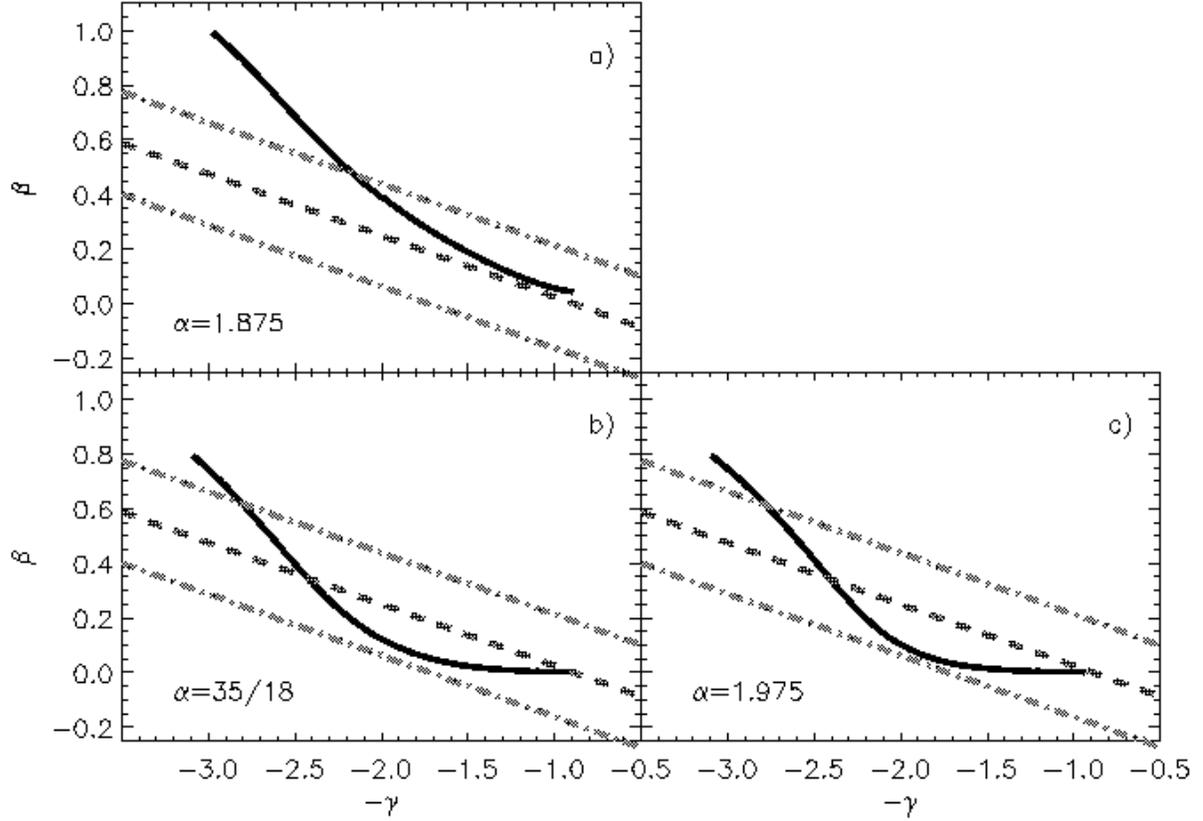}
\figcaption{Correlations between power-law $\beta$
and $-\gamma$ solutions for $10^{-3}x_{\rm vir} \le x \le x_{\rm
vir}$.  Panels a, b, and c represent the best-fit N04 solutions with
$\alpha=1.875,\mbox{ } 35/18, \mbox{ and } 1.975$, respectively.  The
dashed blue and dash-dotted green lines are the same as in
Figure~\ref{plhmnfw}.  The most linear (and hence most qualitatively
consistent with simulations) relationship in panel a is inconsistent
with the quantitative results of simulations.  As in
Figure~\ref{plhmnfw}, the red solution curves that quantitatively
agree with simulations have more concavity than the simulations
themselves.
\label{plhmn04}}
\end{figure}

\begin{figure}
\plotone{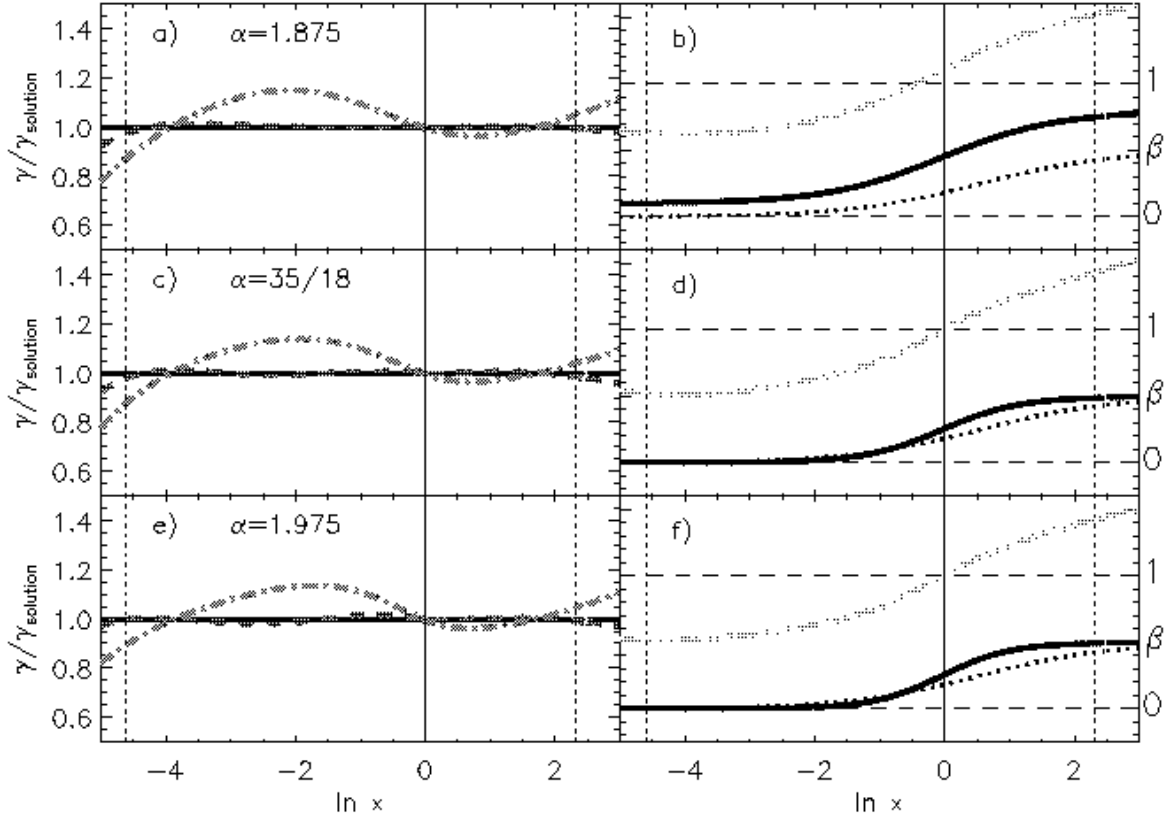}
\figcaption{With tanh $\beta(x)$, panels a, c, and e show the
constrained Jeans equation solution $\gamma$ profiles that best match
the NFW $\gamma$ profile for $\alpha=1.875,\mbox{ }35/18,\mbox{ and
}1.975$, respectively.  Panels b, d, and f illustrate the anisotropy
distributions used to calculate the solutions.  The line styles are
the same as in Figure~\ref{plbnfw}.  Unlike the results shown in
Figure~\ref{plbnfw}, NFW profiles are reproduced very
well by these solutions.  Note that now the $\gamma$ and $\beta$
profiles share the inflection point behavior at $\ln{x}=0$.  This
similarity of $\gamma$ and $\beta$ profile shapes implies a near
linear correlation in the $\beta$--$\gamma$ plane
(Figure~\ref{thmnfw}).  See \S \ref{tanb} for details.
\label{tbnfw}}
\end{figure}

\begin{figure}
\plotone{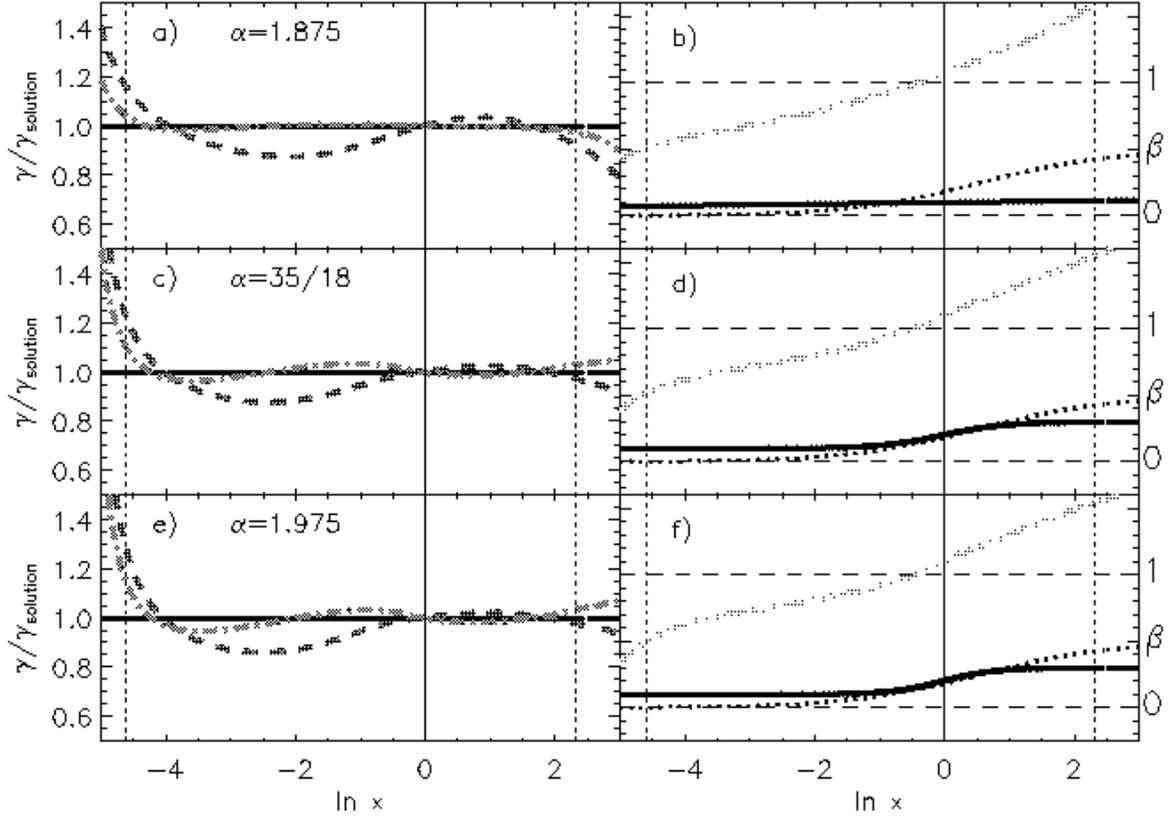}
\figcaption{With tanh $\beta(x)$, panels a, c, and e show the
calculated $\gamma$ profiles that best match the N04 $\gamma$ profile
for $\alpha=1.875,\mbox{ }35/18,\mbox{ and }1.975$, respectively.
Panels b, d, and f illustrate the anisotropy distributions used to
calculate the solutions.  The line styles are the same as in
Figure~\ref{plbnfw}.  These solutions are fair approximations of the
designated N04 profile.  However, note that the $\beta$ profiles are
nearly horizontal.  Higher $\alpha$ values lead to moderate variations
in anisotropy, but not as large as those in Figure~\ref{tbnfw}.
\label{tbn04}}
\end{figure}

\begin{figure}
\plotone{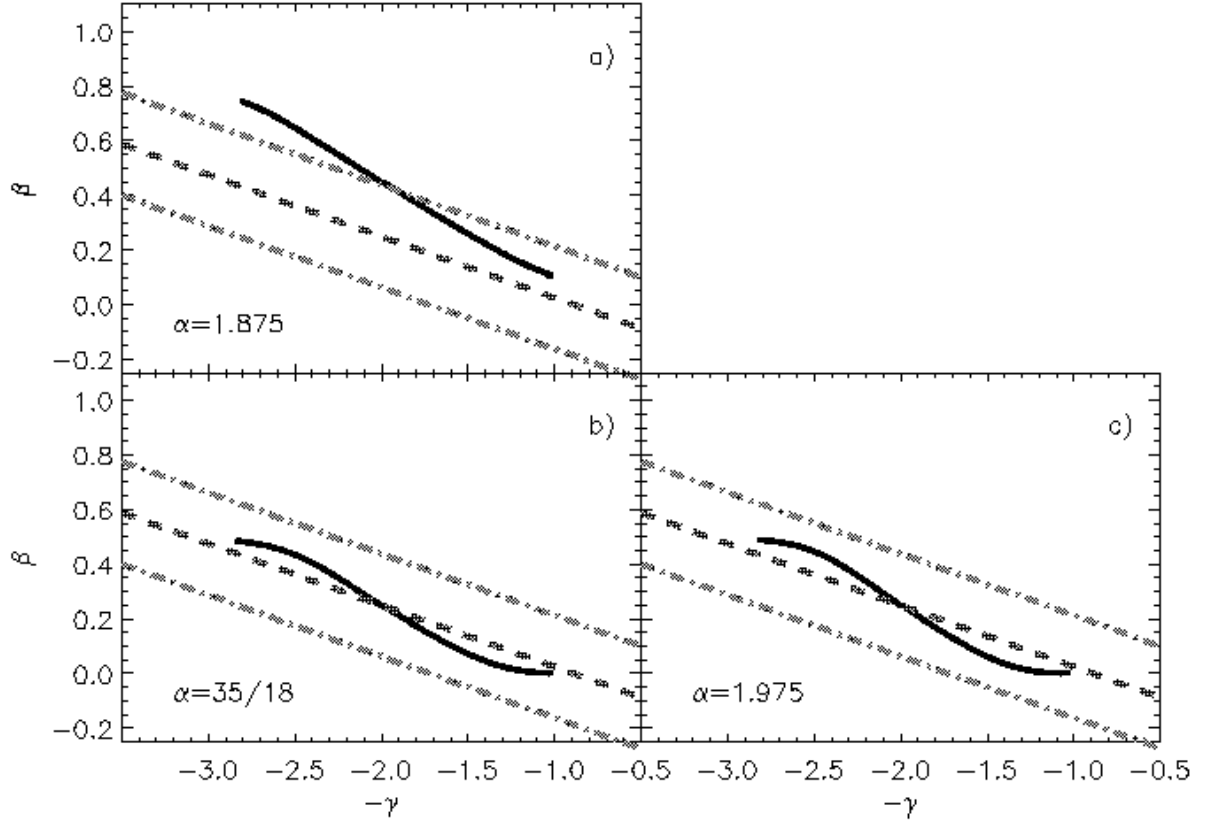}
\figcaption{Correlations between tanh $\beta$ and
$-\gamma$ for $10^{-3}x_{\rm vir} \le x \le x_{\rm vir}$.  Panels a,
b, and c represent the best-fit NFW solutions with
$\alpha=1.875,\mbox{ } 35/18, \mbox{ and } 1.975$, respectively.  The
dashed blue and dash-dotted green lines are the same as in
Figure~\ref{plhmnfw}.  As in Figure~\ref{plhmn04}, the most linear
relation is quantitatively inconsistent with the results of
simulations.  Higher $\alpha$ values produce red curves that have
larger regions of linearity and are consistent with simulations.
Recall that in this case the $\gamma$ and $\beta$ profiles have
similar shapes (particularly the inflection point at $\ln{x}=0$).
\label{thmnfw}}
\end{figure}

\begin{figure}
\plotone{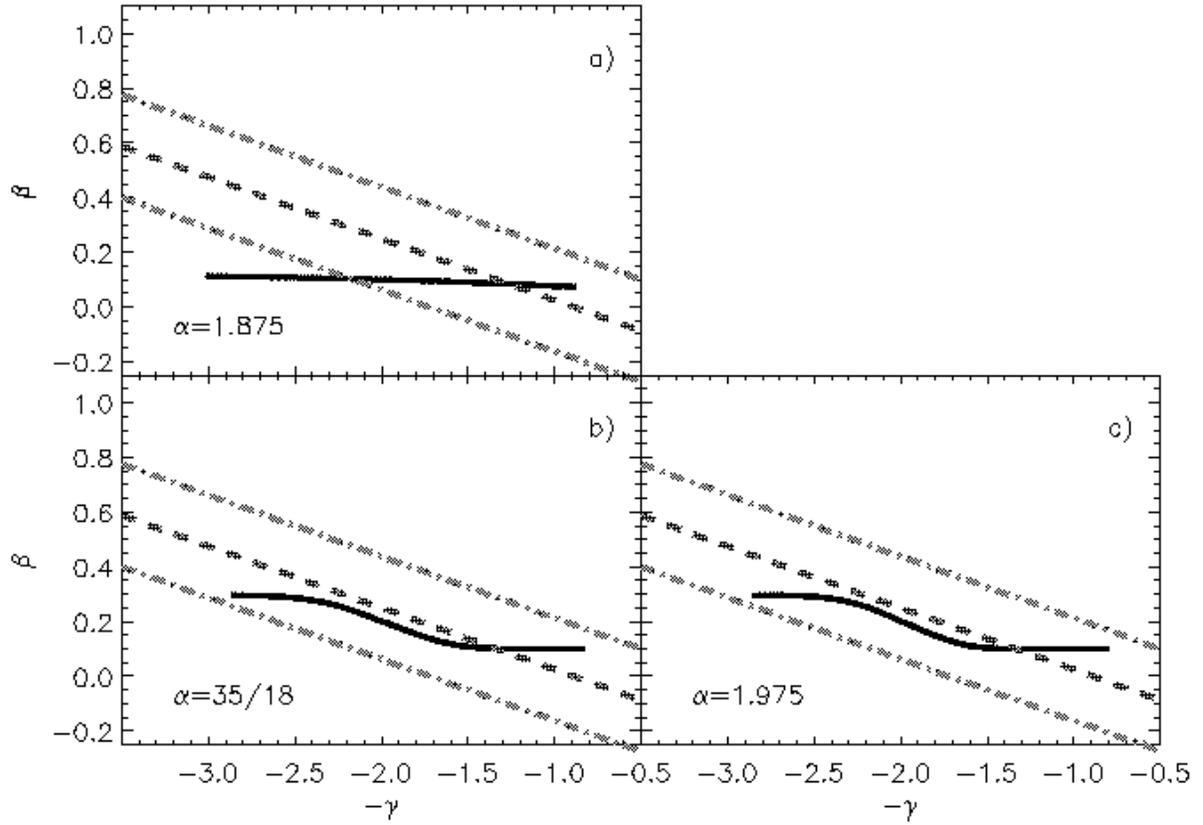}
\figcaption{Correlations between tanh $\beta$ and
$-\gamma$ for $10^{-3}x_{\rm vir} \le x \le x_{\rm vir}$.  Panels a,
b, and c represent the best-fit N04 solutions with
$\alpha=1.875,\mbox{ } 35/18, \mbox{ and } 1.975$, respectively.  The
dashed blue and dash-dotted green lines are the same as in
Figure~\ref{plhmnfw}.  Overall, the red curves are more horizontal
than the simulation results.  However, larger $\alpha$ values allow
for quantitative agreement between the solution curves and
simulations.
\label{thmn04}}
\end{figure}

\begin{figure}
\scalebox{0.8}{
\plotone{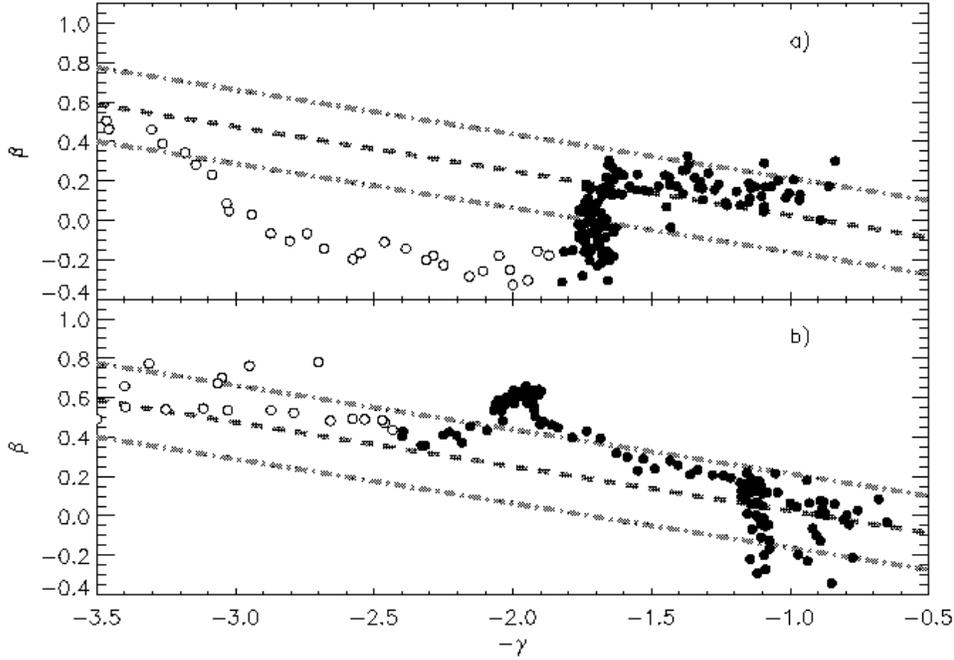}}
\figcaption{The $\beta$ vs. $-\gamma$ curve for the standard ESIM halo
(a).  The corresponding curve for the ESIM halo that has been modified
so that the density profile resembles that of an NFW model.  The
points in each panel marked by red solid circles represent parts of
the halo that lie in the range $10^{-3} < r/r_{200} < 0$, where
$r_{200}$ is the virial radius of those simulations.  The open circles
lie beyond the virial radius.  The dashed blue and dash-dotted green
are the same N-body simulation results shown in the previous plots of
$\beta$ vs. $\gamma$.  Despite the bump at $\gamma=2$, panel b
resembles the results from N-body simulations much more closely than
does panel a.  Each of these ESIM halos has power-law \psd profiles,
but they have radically different $\gamma$--$\beta$ relationships.
This suggests that the mechanisms that give rise to each of these
relations are independent.  The standard halo (panel a) is nearly
isotropic out to the virial radius while the NFW-like halo (panel b)
has a tanh-like $\beta$ profile.  This prompts us to postulate that
the $\beta$--$\gamma$ connection is rooted in the instability that
creates isotropic cores and radially anisotropic outer regions, and
that the \psd\ relationship is due to a more general physical process.
\label{esim}}
\end{figure}

\end{document}